\documentclass{mem}
\usepackage{natbib}
\usepackage{txfonts}
\usepackage{graphicx}
\idline{0}{1}
\bibpunct{(}{)}{;}{a}{}{,} 

\begin{document}
\def\teff{$T\rm_{eff }$}
\def\kms{$\mathrm {km s}^{-1}$}

\title{
Abundances of stars in different Galactic subsystems
}
\subtitle{}
\author{Thomas Bensby} 
\institute{Lund Observatory, Department of Astronomy and Theoretical 
Physics, Box 43, SE-221\,00 Lund, Sweden, \email{tbensby@astro.lu.se}}
\authorrunning{T.\,Bensby}
\titlerunning{Abundances of stars in different Galactic subsystems}

\abstract{
This is a brief overview of the elemental abundance patterns that have
been observed in the different Galactic stellar populations. 
Main focus is on studies that are based on
high-resolution spectra of dwarf and subgiant stars, and in some cases
red giant stars. Of particular interest is the thin-thick disk dichotomy, 
the variation of abundances and stellar ages with galactocentric radius, 
multiple stellar populations in the Galactic bulge region, and how some of 
these may be connected with other Galactic populations.
\keywords{Stars: abundances --
Stars: atmospheres -- Stars: Population II -- Galaxy: globular clusters -- 
Galaxy: abundances -- Cosmology: observations }
}

\maketitle{}

\section{Introduction}

The current understanding of our Milky Way galaxy is that it is a grand spiral
galaxy containing a multi-population boxy/peanut-shaped bulge \citep{lopezcorredoira2005,ness2013}, a dual disk system 
\citep{gilmore1983}, and a dual stellar halo
\citep{carollo2007}. Furthermore, there are stellar
overdensities and structures in velocity space 
\citep[e.g.,][]{antoja2012} as well as in 
three-dimensional space \citep[e.g.,][]{belokurov2006}. This galactic complexity
with different stellar subsystems and different origins hold
the secrets and pieces to puzzle on galaxy formation and evolution that 
slowly is being laid.

It is in recent years that this complex picture of the main Galactic 
stellar populations (disk, bulge, and halo) has emerged. First, based on star 
counts towards the Galactic South Pole \cite{gilmore1983} found the
that the galactic disk contains two disk populations, a thin disk
and a thick disk. The thick disk has since then been observed to contain stars
that on average are older, more metal-poor, and, at a given metallicity, 
more $\alpha$-enhanced than the stars of the thin disk. These differences
point to that they are distinct stellar populations with different origins.
However, \cite{bovy2012} claimed that there is no distinct 
thick disk in the Milky Way, and that the two disks form a continuous sequence 
of mono-abundance populations with increasing scale-heights. 
Contradictory, \cite{lee2011} and \cite{liu2012} utilised
the same data set as used by \cite{bovy2012} and 
found two or perhaps even three distinct components of the Galactic disk. 
As the existence, or non-existence, of distinct thick disks in spiral disks 
galaxies is an important signature of the galaxy formation scenario 
\citep[e.g.,][]{minchev2012}, it is crucial to firmly establish the properties
of the Galactic disk. Next, the picture of the bulge has changed. 
A decade ago it was generally believed that the bulge was a classical
bulge, originating in the early collapse of the pre-galactic cloud
\citep{eggen1962}, and hence containing
only old and metal-rich stars \citep[see, e.g., review by][]{wyse1997}. 
Now, most observational evidence indicate that the bulge was formed from
disk material, i.e. it is a pseudo-bulge \citep[see review by][]{kormendy2004}, 
and it has been found to contain a significant fraction of young and intermediate
age stars. Lastly, the status of the stellar halo has also changed as well. 
It now appears as if there is at least two halos, one inner halo and one 
outer halo, with different chemical as well as kinematical properties
\citep[see][]{nissen1997,nissen2010,carollo2007}.  

To trace the origins of these structures and stellar subsystems in the Milky Way
it is utterly important to have clear pictures of their age, elemental 
abundance, and velocity distributions. However, kinematic properties may 
change with time due to gravitational interactions with other stars, gas clouds, and/or spiral arms. Also, a mechanism
called radial migration \cite{sellwood2002} has in recent years gained a lot 
of interest, and if real, stars may ``ride'' on spiral waves and change from 
one circular orbit to another, and hence destroying the kinematical signatures 
of its formation. Instead, a stellar property that remain untouched 
through the history of the Galaxy is the atmospheric chemical composition of 
low-mass stars. In particular, relatively un-evolved F and G type 
stars on the main sequence, turn-off, or subgiant branch have lifetimes 
exceeding the age of the galaxy, during which their atmospheres remain intact
and trace the chemical composition of the gas clouds they formed from.
These types of stars have therefore been extensively used during the last 
decades to map the abundance structure of the disk and halo in the Solar 
neighbourhood \citep{edvardsson1993,fuhrmann1998,reddy2003,bensby2003}. 
These types of stars have, however, a major drawback, 
and that is that they are intrinsically faint and cannot under normal 
circumstances be studied at large distances with high-resolution 
spectrographs.  In order to study the abundance distribution at larger
scales in the Galaxy one therefore turn to brighter stars, such as
red giant stars.

I current era of large scale spectroscopic surveys, aiming at mapping the
Galactic disk, halo, and even the bulge, several kpc from the Sun 
with hundreds of thousands of stars, it is of course impossible
to constrain the star samples to un-evolved stars dwarf stars. Surveys
such as GALAH \citep{zucker2012} and APOGEE \citep{allendeprieto2008} 
therefore mainly target red giants stars and aim
to obtain relatively high $S/N$ spectra. The Gaia-ESO survey \citep{gilmore2012}, 
on the other hand,
is mainly targeting the turn-off region, and will have to settle with the fact
that the obtained spectra can have somewhat lower $S/N$. This will present 
a challenge to the analysis, that has to be carefully done and anchored, using
stars with known properties, so called benchmark stars. It is also important
to have good comparison samples, tracing different Galactic populations,
analysed from high-resolution and high signal-to-noise spectra.
In this paper I will give a brief overview of the abundance structure
observed in the Galactic disk and bulge.

\section{The solar neighbourhood disk}

The part of the Galaxy that has been best mapped is the Solar neighbourhood.
Following the seminal study by \cite{edvardsson1993}, \cite{fuhrmann1998}
presented the first in a series of papers aimed at producing an unbiased 
volume-complete sample for all mid F-type to early K-type stars down to absolute
magnitude $M_V=6$, north of declination $-15^\circ$, within a radius of 
25\,pc from the Sun. Thirteen years later this sample was 85\,\% complete
and contained more than 300 solar type stars \citep{fuhrmann2011}.
What Fuhrmann found was that that the sample divided roughly into two types
of stars; one with stars that were old and that had high [Mg/Fe]
abundance ratios, and one where the stars where young and that had low
[Mg/Fe] abundance ratios. Most of the old stars were classified as
thick disk stars while the young stars were classified as thin disk stars.
In the [Mg/Fe]-[Fe/H] plane these two groups showed extremely well-defined
and distinct abundance trends, with the thick disk stars laying on an elevated
and flat [Mg/Fe] plateau ranging from $\rm [Fe/H]\approx -0.9$ up to
$\rm [Fe/H]\approx -0.25$. The thin disk abundance trends were offset
from the thick disk trends and showed a shallow decline in 
[Mg/Fe] from $\rm [Fe/H]\approx -0.6$ up to $\rm [Fe/H]\approx +0.4$.
Only 15 of the more than 300 stars in the Fuhrmann studies were classified
as thick disk stars, and it is clear that in order to really probe the
abundance trends of the Galactic thick disk one has to extend to slightly larger
distances. 

The best strategy would of course be to observe thick disk stars where
the thick disk is the dominating population. With current estimates of 
scale-heights and stellar density normalisations in the plane, one has to go
distances of 1-2\,kpc from the plane in order to obtain a sample
that is dominated by thick disk stars. However, at these distances F and G 
dwarf stars have apparent magnitudes fainter than $V\approx15$ and it is 
no longer feasible to obtain high-resolution and high signal-to-noise spectra 
for large samples. Instead, a convenient way to identify potential thick disk 
candidates for high-resolution spectroscopic observations
is through kinematical selection criteria. These relies on the assumption
that the velocity distributions of the different stellar populations have
Gaussian distributions, that the average rotation velocities around the Galactic 
centre are different, and that they occupy certain fractions of the stellar density
in the Galactic plane. Applying the kinematic criteria defined in \cite{bensby2003}
to the 16\,000 stars in the Geneva-Copenhagen survey (GCS) by \cite{nordstrom2004} 
around 500 stars have kinematic properties classifying them
as candidate thick disk stars. Many of those stars have metallicities
well above solar values \citep[see, e.g.,][]{bensby2013disk}.

\begin{figure*}
\centering
\resizebox{\hsize}{!}{
\includegraphics[bb= 0 160 445 620,clip]{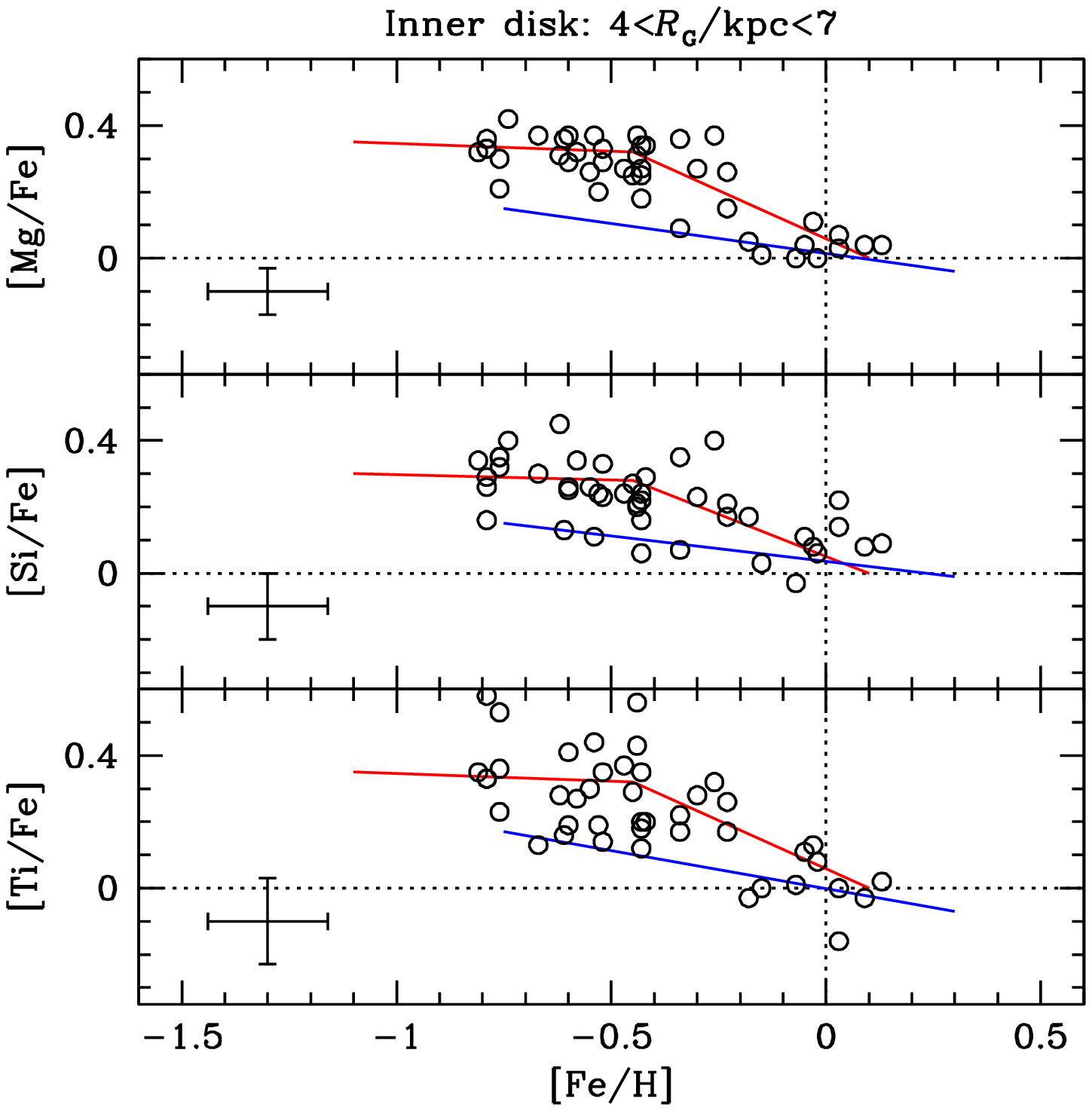}
\includegraphics[bb=75 160 445 620,clip]{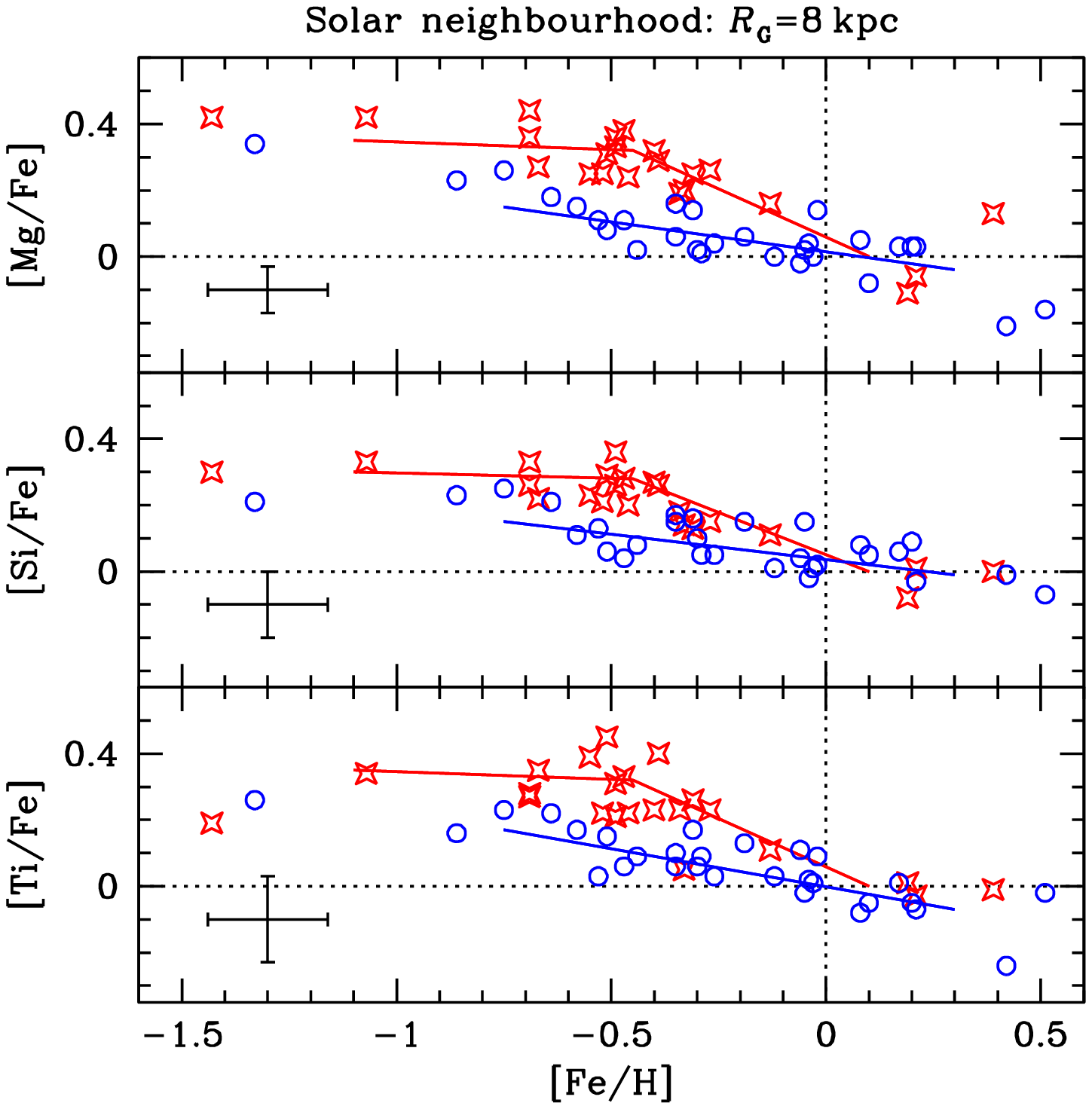}
\includegraphics[bb=75 160 465 620,clip]{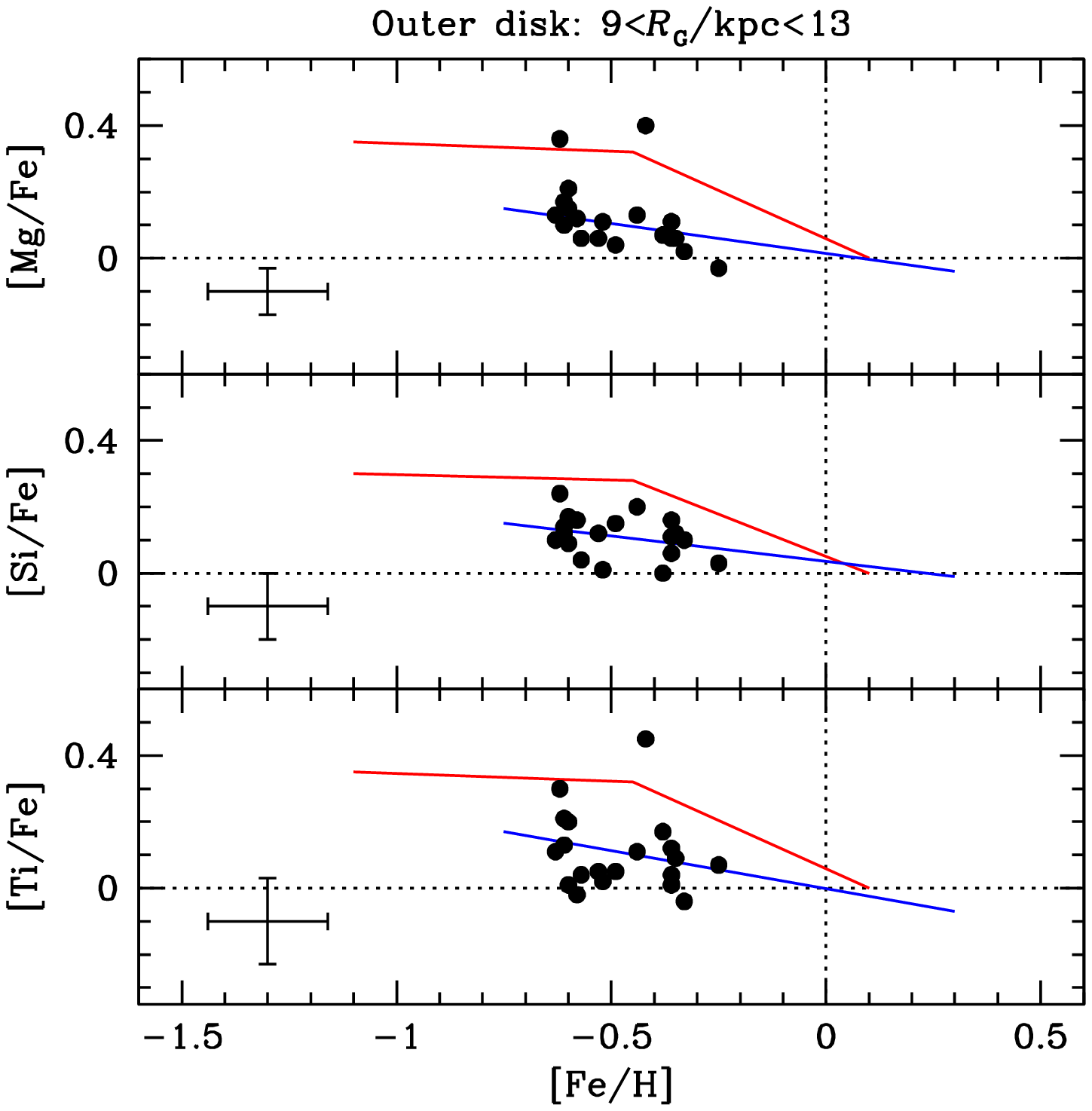}}
\caption{
Abundance trends for the $\alpha$-elements Mg, Si, and Ti.
The left panel shows the 44 inner disk red giants from
\cite{bensby2010letter}, the centre panel shows the solar neighbourhood
thin and thick disk stars
(blue circles and red stars, respectively) by \cite{alvesbrito2010}.
The right panel shows the 20
outer disk red giants from \cite{bensby2011letter}.  
The red and blue lines in the abundance
plots are fiducial lines based on the solar neighbourhood abundance
trends. Figure is taken from \cite{bensby2011letter}.  
\label{fig:innerouterdisk}
}
\end{figure*}

\cite{feltzing2003} and \cite{bensby2003} showed that kinematically
selected thick disk stars most likely reach higher [Fe/H] than observed by the
\cite{fuhrmann1998} and that the signature of chemical enrichment
from supernovae type Ia shows up as a downturn (the ``knee'') in the thick disk
$\alpha$-to-iron abundance trends. Other studies have also utilised kinematical 
criteria to define thick disk star samples and find similar results
\citep[e.g.,][]{reddy2003,reddy2006,soubiran2003,bensby2004,bensby2005},
and \cite{bensby2007letter2} showed that the thick disk
reaches at least solar metallicities. A similar dichotomy has also been found
by \cite{adibekyan2012} that analysed the HARPS sample that initially was
observed to detect exoplanets. The most recent study is \cite{bensby2013disk}
who has selected a sample of 714 nearby F and G dwarf and subgiant 
in order to trace the extremes of the different Galactic stellar populations:
the metal-poor limit of the thin disk, the metal-rich and metal-poor
limits of the thick disk, the metal-rich limit of the halo, structures in 
velocity space such as the Hercules stream and the Arcturus moving group, 
and stars that have kinematical properties placing them in between the thin 
and thick disks. Among other things they show that kinematical selection criteria
produce a lot of mixing between selected samples. A kinematically selected thick 
disk sample is polluted by high-velocity thin disk stars, and a kinematically
selected thin disk sample is polluted by low-velocity thick disk stars.
A better discriminator appears to be stellar ages, and a good dividing age
between the thin and thick disk appears to be around 8\,Gyr (see also
\citealt{haywood2013}). Unfortunately, as stellar ages at high precision
are notoriously difficult to estimate, the large age uncertainties is likely
to produce a mixing between age-selected thin and thick disk samples of the same
magnitude as between kinematically selected samples. 
\cite{bensby2013disk} also show that the distinction between the thin
and thick disk abundance trends becomes much clearer if stars that are 
coupled to higher uncertainties in the abundance ratios are rejected.

In summary, high-resolution and high signal-to-noise spectroscopic studies
of solar-type dwarf stars in the Solar neighbourhood shows an abundance
dichotomy in the Galactic disk, and that this dichotomy is associated with two 
different stellar populations that appear to be separated in time. The 
question is if this dichotomy persists if we move away from the immediate solar 
neighbourhood?

\section{Inner and outer disk}

Beyond the Solar neighbourhood the Galactic disk is sparsely 
studied. There are some studies in the inner disk
of young bright O and B type stars 
\citep[e.g.,][]{daflon2004} and Cepheids \citep[e.g.,][]{luck2006},
that both trace the chemical composition of the present day Galactic disk. 
The outer disk has been somewhat better studied using stars 
in open clusters \citep[e.g.,][]{yong2005,carraro2007,jacobson2011}, 
O and B type stars \citep[e.g.,][]{daflon2004,daflon2004outer}, as well 
as Cepheids \citep[e.g.,][]{andrievsky2004,yong2006}, which all are 
tracers of relatively young populations. Field red giants, that can 
trace older populations, were observed by \cite{carney2005}, but
their sample only contained three stars.

\begin{figure*}
\centering
\resizebox{0.65\hsize}{!}{
\includegraphics{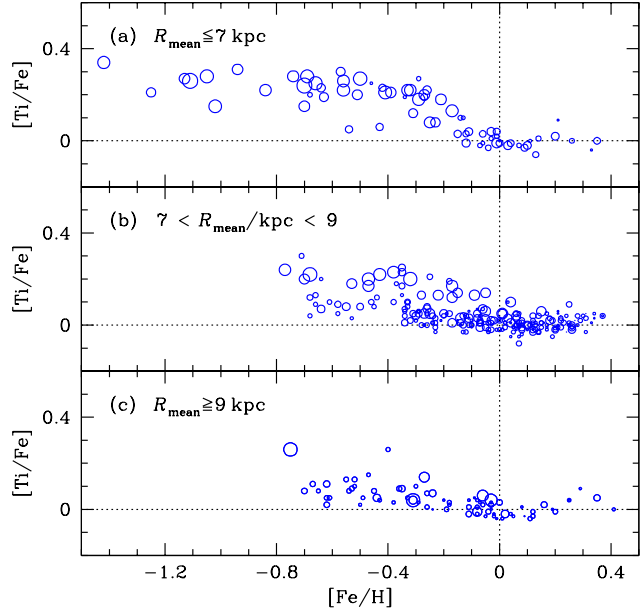}}
\caption{[Ti/Fe]-[Fe/H] abundance trends
for stars with different $R_{\rm mean}$.
Only stars for which he difference between the upper and
lower age estimates is less than 4\,Gyr are included.
The sizes of the circles have been scaled
with the ages of the stars. Figure is taken from \cite{bensby2013disk}.
\label{fig:rmeantife}
}
\end{figure*}

The situation was improved with the studies
by \cite{bensby2010letter} and \cite{bensby2011letter} that presented detailed 
elemental abundances in 44 red giants in the inner disk and 20 red giants in 
the outer disk. The inner disk sample spans galactocentric radii 3 to 7\,kpc, 
and the outer disk sample galactocentric radii 9 to 12\,kpc. The stars of both 
the inner and outer disk samples are located up to 3 to 4\,kpc from the plane 
and should trace both the thin and thick disk abundance trends if present 
at these locations in the Galaxy.

Figure~\ref{fig:innerouterdisk} shows a comparison of the abundance trends as 
traced by the in situ red giant stars in the inner and outer disks from 
\cite{bensby2010letter} and \cite{bensby2011letter}, to the local thin and 
thick disk abundance trends, as traced by red giant stars, from 
\cite{alvesbrito2010}. First, it is evident that the local red giant comparison 
sample shows the characteristic thin and thick disk abundance trends as seen
by the local dwarf studies cited in the previous section. 
In the left panels of Fig.~\ref{fig:innerouterdisk} we then see that
the inner disk red giant sample occupy the region in abundance space where 
thin and thick disk stars are found in the solar neighbourhood, however
with a deficiency of stars in the metal-poor ``thin disk abundance region'' 
below $\rm [Fe/H]\approx -0.2$. The outer disk red 
giant sample, on the other hand, shows a completely different abundance 
pattern. Essentially no one of the 20 outer disk stars, which are
located up to 2\,kpc from the plane (and one even 6\,kpc from the plane), 
seem to have an abundance pattern 
resembling what is seen in the local thick disk. Instead they occupy a rather 
narrow metallicity regime around $\rm [Fe/H]\approx -0.5$, and none (or maybe 
one or two for Mg and Ti) show elevated $\alpha$-to-iron abundance ratios. 
\cite{bensby2011letter} interpreted the lack of ``thick disk stars'' in the 
outer disk as being due to that the thick disk scale-length is shorter than 
that of the thin disk. \cite{juric2008} had previously estimated the 
thick disk scale-length to be 3.6\,kpc and that of the thin disk to be 2.6\,kpc. 
New estimates from \cite{bensby2011letter} was 2.0\,kpc for the thick disk, 
and 3.8\,kpc for the thin disk, i.e., quite the opposite behaviour, meaning 
that the thick disk stellar density drops of much more quickly with galactocentric
radius than the thin disk stellar density.  The 
short scale-length for the thick disk has later been confirmed by 
\cite{cheng2012_2} who based on the SDSS Segue G dwarf sample found scale-lengths 
of 3.4\,kpc and 1.8\,kpc for the thin and thick disks, respectively.
A short thick disk scale-length is now also required in dynamical models
of the Galaxy to match the observations \citep[e.g.,][]{binney2013}.

The lack of $\alpha$-enhanced stars in the outer disk is also evident from the
local sample of 700 F and G dwarf stars from \cite{bensby2013disk}. 
Figure~\ref{fig:rmeantife} shows the [Ti/Fe]-[Fe/H] trends for three
subsamples, separated into three bins of the mean orbital galactocentric 
distance ($R_{\rm mean}\equiv (R_{\rm min} + R_{\rm max}) / 2$) and it is clear that old and $\alpha$-enhanced stars with $R_{\rm mean}>9$\,kpc is lacking.
We furthermore note that results from the first year
of APOGEE data confirm the lack of $\alpha$-enhanced stars in the outer 
disk \citep{anders2013}.

In summary, the thin and thick disk dichotomy as observed 
in the solar neighbourhood is also present in the inner disk, 
although with a different relative contributions than observed
at the solar radius. At the solar radius,
in the Galactic plane the thick disk is around 10\,\%, but as a result of 
the short scale-length it increases when going to shorter galactocentric radii 
and decreases when going to larger radii. In a way, the thick disk 
appears to be truncated at, or slightly beyond, the solar radius. 

\begin{figure*}
\centering
\resizebox{\hsize}{!}{
\includegraphics[bb=18 144 592 718,clip]{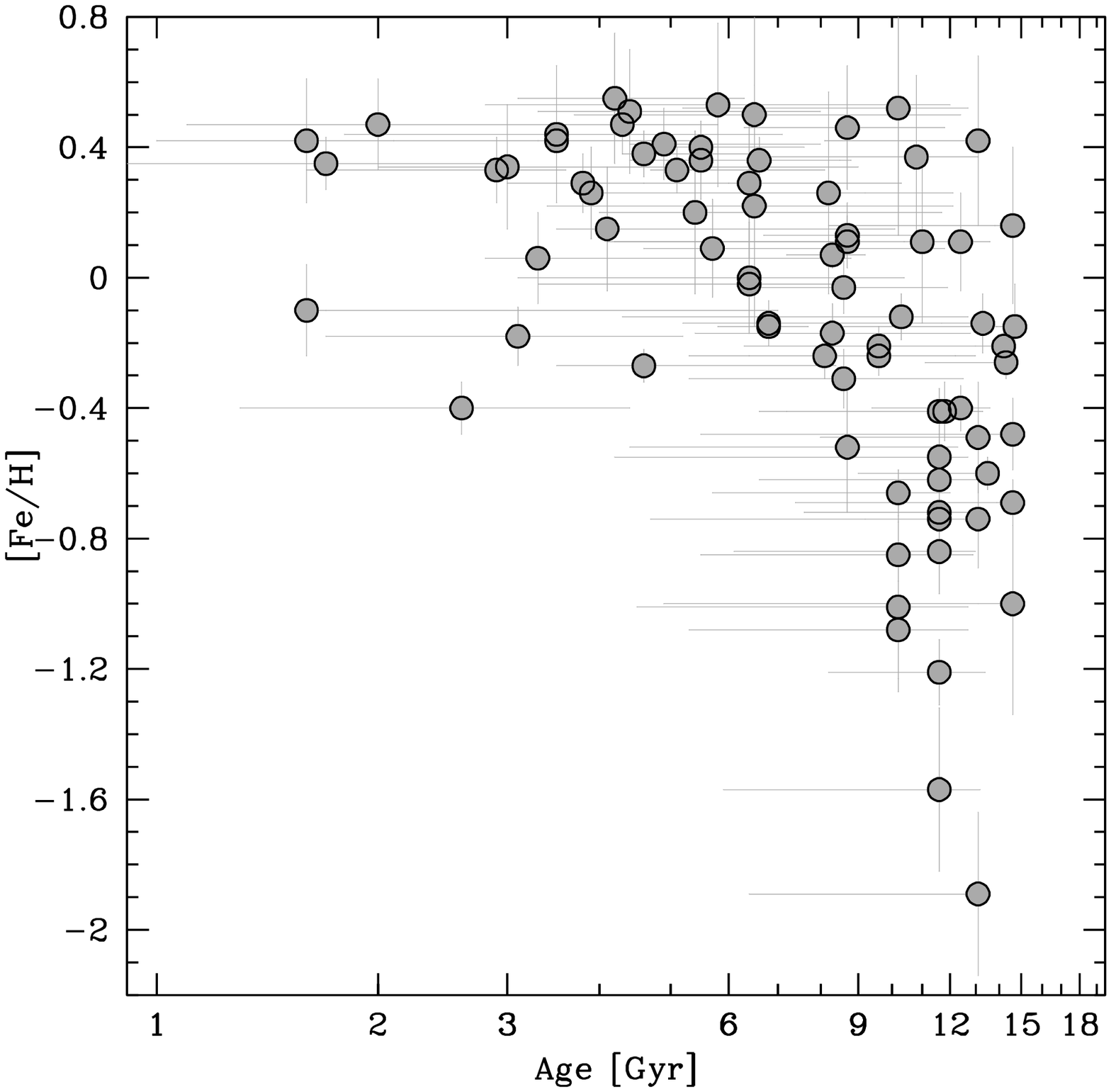}
\includegraphics[bb=18 144 592 718,clip]{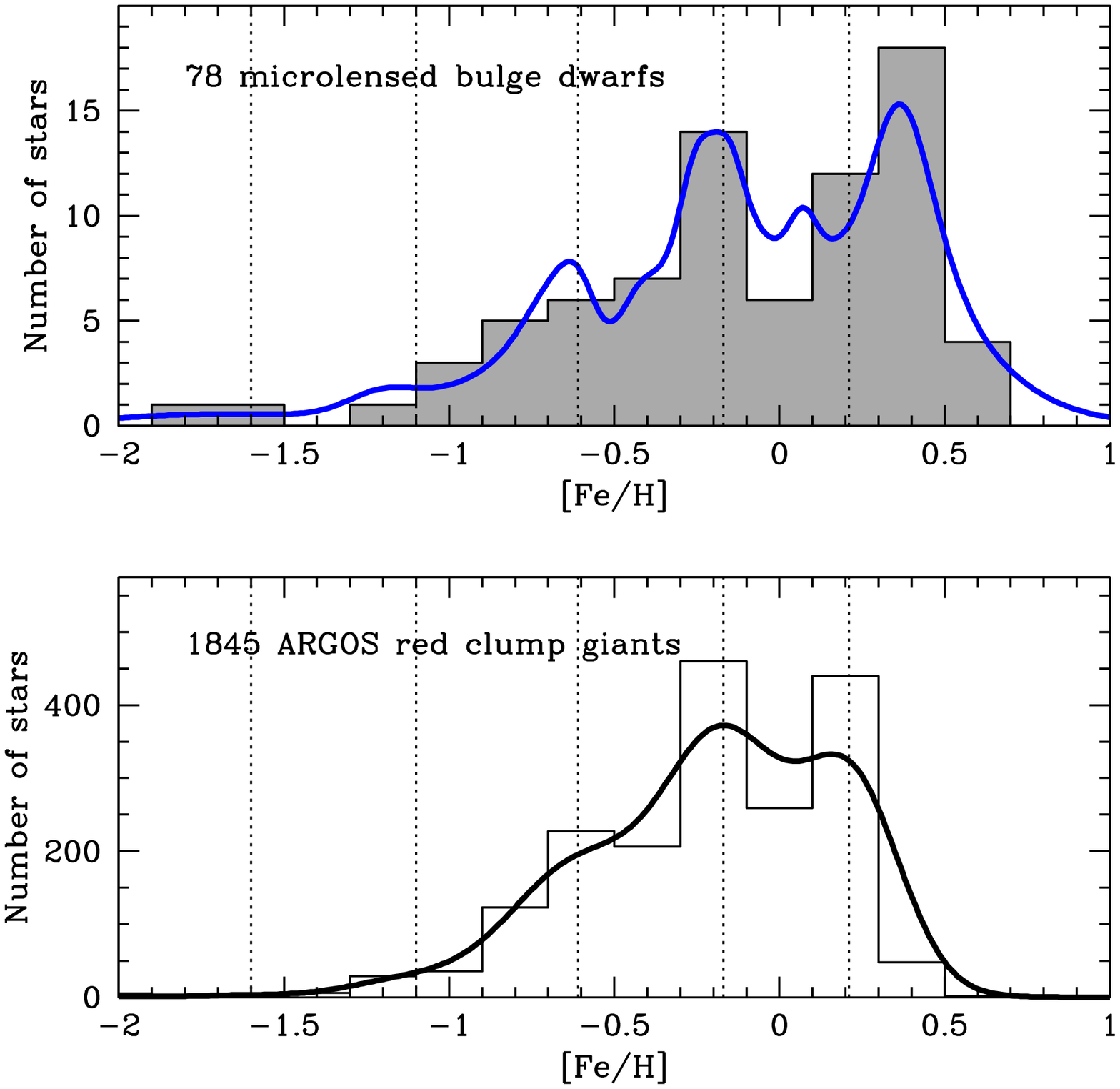}}
\caption{
Left-hand plot shows the age-metallicity diagram for the now in total
78 microlensed bulge dwarf and subgiant stars (58 microlensed dwarf
stars from \cite{bensby2013}, as well as 20 new
stars from Bensby et al.~(2014, in prep.)).
Right-hand plot shows the metallicity distribution for the microlensed
dwarf stars. Also shown is the MDF for 1845 red 
giant stars from the ARGOS survey fields at 
$\rm (l, b) = (0, 5), (5, -5), (-5, -5)$ 
from \cite{ness2013}. The curved lines represent generalised histograms,
and the dotted vertical lines mark the peaks claimed by \cite{ness2013}.\label{fig:bulge}
}
\end{figure*}
\section{Bulge}

Since the study by \cite{mcwilliam1994} who presented the first detailed 
elemental abundance study of red giants in Baade's window, the chemical 
evolution of the bulge has been extensively studied using red giants 
\citep[e.g.][]{zoccali2006,fulbright2007,lecureur2007}. 
These studies all showed that the $\alpha$-to-iron abundance ratios were 
elevated even at super-solar metallicities, meaning that the bulge had 
experienced a very fast chemical enrichment, much faster than in 
any of the other stellar populations. As most of the large Solar 
neighbourhood studies had utilised dwarf stars comparisons had to be 
made between disk dwarfs and bulge giants. This changed with 
\cite{melendez2008}
who did a differential analysis between bulge giants (including a re-analysis
of the \citealt{fulbright2007} sample) and local thin and thick disk giants. 
They did not find the high $\alpha$-element abundance
levels at super-solar metallicities, but instead that the bulge abundance trends 
compared very well with what was observed in local thick disk red giants.
Recent studies of red giant star samples in the bulge confirm
these results \citep[e.g.,][]{babusiaux2010,ryde2010,hill2011,gonzalez2011}.

A new window to study the abundance structure of the Galactic bulge
opened with the advent of microlensing surveys such as OGLE \citep{udalski1994}
and MOA \citep{bond2001} that both have early warning systems that
alert when new microlensing events occur, and when they will reach peak brightness. 
If the un-lensed magnitude of the background source is consistent with a
dwarf star located at a distance of 8\,kpc in the bulge, and if the
peak magnification is sufficient, it is possible to obtain a high signal-to-noise
high-resolution spectrum with 1-2 hours integration with the current generation
of 8-10\,m telescopes. The first detailed abundance analysis of a
dwarf star in the bulge based on a high-resolution spectrum obtained 
during a microlensing event 
was presented by \cite{johnson2006}. Since then we have conducted an 
intense observing campaign, mainly through a target-of-opportunity program 
with UVES at VLT. 
Adding 20 new targets from the 2013 season to the 58 targets published
in \cite{bensby2013} the sample currently consists of 78 dwarf and subgiants 
in the bulge (Bensby et al., 2014, in prep).

A comparison between the 58 bulge dwarfs in \cite{bensby2013} 
and the 700 nearby dwarf stars in \cite{bensby2013disk} confirms the similarities
between the bulge and thick disk abundance trends. There are indications that 
position of the knee in the bulge $\alpha$-to-iron abundance trends occur at 
a slightly higher metallicity (about 0.05\,dex) than for the thick disk, 
indicating the possibility that
the bulge experienced a slightly faster chemical enrichment history
\citep{bensby2013}. 

With the microlensed bulge dwarf it has also
been possible to determine ages for individual stars in the bulge.
The age-metallicity
diagram in Fig.~\ref{fig:bulge}, containing all to date 78 microlensing events,
shows that the bulge contain a significant fraction of young 
and intermediate age stars. This is contradiction with the old bulge claimed by 
studies of the colour magnitude diagram in different bulge fields
\citep[e.g.,][]{zoccali2003,clarkson2008}.
As was shown in
\cite{bensby2013} a metal-poor and old isochrone is very hard to distinguish
from a metal-rich and young isochrone. If the metallicities of individual
stars are not know, which is the case in most photometric studies,
the CMDs will show an apparent old turn-off. 

In addition, results from the ARGOS survey claim multiple components
in the bulge metallicity distribution \citep{ness2013}. 
Now when statistics for the microlensed dwarfs
are starting to build up, the components claimed by \cite{ness2013} are 
evident also there (see Fig.~\ref{fig:bulge}). 
In combination with the results from the BRAVA survey,
showing that the bulge has cylindrical rotation \citep{kunder2012}, 
there is very little room left for a classical collapse population in the 
Galactic bulge. Instead it appears as if the bulge is the central region of 
the Milky Way where the other Galactic stellar populations meet and overlap.

\section{Outlook}

The ongoing large spectroscopic surveys (e.g., Gaia-ESO, GALAH, and
APOGEE) in conjunction
with astrometric data from the upcoming Gaia satellite
will provide a gold mine
for Galactic archaeology and our understanding for the origin
and evolution of the Milky Way.

\begin{acknowledgements}
T.B. was funded by grant No. 621-2009-3911 from The Swedish 
Research Council.
\end{acknowledgements}

\bibliographystyle{aa}
\bibliography{referenser}

\end{document}